\documentclass{epl}
\usepackage{epsfig}
\newcommand{\suu}{$\mathrm{SU(3)_c} \times
\mathrm{SU(3)_L} \times \mathrm{U(1)_X}$}
\newcommand{\tto}{{\bf 3-3-1} model}

\title{Photon-photon scattering in a  {\bf{3-3-1}}  model}

\author{G. Tavares--Velasco \inst{1} \and J.J. Toscano \inst{2}}

\institute{ \inst{1} Departamento de F\'{\i}sica, CINVESTAV, Apartado Postal
 14--740, 07000, M\'exico D.F., M\'exico \\
  \inst{2}Facultad de Ciencias
  F\'{\i}sico Matem\'aticas, Universidad Aut\'onoma de Puebla, Apartado
  Postal 1152, Puebla, Pue., M\'exico 
} 

\pacs{4.80.-j}{Other particles (including hypothetical)}

\begin{document}

\maketitle

\abstract{ We analyze the effects of a doubly charged vector bilepton as
well as exotic quarks with charge $5/3~e$ and $-4/3~e$ on light by light
scattering. We consider mass values in the range 0.3--1 TeV, which would
be reached at the planned future linear colliders. It is found that such
exotic particles, especially the doubly charged vector bilepton, give
raise to remarkable deviations from the standard model cross section. The
virtual effects arising from these particles would provide an indirect
test to a particular model which is based on the \suu~ gauge symmetry,
known as \tto, where such particles are a natural prediction.}

\section{Introduction}

In the past, light by light scattering was the subject of theoretical
interest in an early attempt of applying perturbation theory beyond the
tree level \cite{Euler}. The cross section has been studied to a great
extent \cite{Jikia}, but owing to technical troubles its experimental
scrutiny has deserved little attention. Nevertheless, it is likely that
$e^+e^-$ linear colliders (LC) would be built in the future
\cite{LinColl}. This class of colliders would also operate in the $\gamma
\gamma$ and $\gamma e$ modes, opening up the possibility of performing a
series of interesting experiments which are inaccessible at lepton and
hadron colliders. The prospect of the construction of LC has renewed the
interest in light by light scattering: it was found recently that this
reaction might be an efficient mode to search for virtual effects of
particles lying beyond the standard model (SM) \cite{Renard}. Following
this approach, the virtual effects produced by particles which arise in
supersymmetry (SUSY), left-right models (LRM), and supersymmetric
left-right models (SUSYLR) have been examined in \cite{Renard,Toscano1}.
In this letter we will consider the inclusion of a doubly charged gauge
boson as well as exotic quarks with charge $5/3~e$ and $-4/3~e$. Although
the analysis is applicable to any model predicting such exotic particles,
the main motivation lies in an extension of the SM based on the \suu~
gauge symmetry, known as \tto, where these particles appear naturally
\cite{Pisano1}.

\section{A $3-3-1$ model}

The idea of extending the standard model (SM) by embedding the electroweak
gauge group into $\mathrm{SU(3)_L}\times \mathrm{U(1)_X}$ is previous to
the appearance of the \tto. One of the motivations of this model is the
necessity of having a chiral theory of bilepton gauge bosons.
\footnote{Non-gauge vector bileptons may appear in composite and
technicolor theories.} These exotic particles arise also from an
$\mathrm{SU(15)}$ grand unified theory with proton-stability, which has
however the drawback of requiring artificial mirror fermions in order to
cancel anomalies \cite{GUT}. On the contrary, the \tto~ is appealing for
suggesting a possible path to the solution of the flavor problem: anomaly
cancellation is achieved only if all the fermion families are added up,
instead of the usual cancellation occurring when it is summed over each
fermion family separately. As a consequence, the number of fermion
families must be a multiple of 3, the number of quark colors. In addition
to this peculiarity, this model is interesting because it predicts
striking effects which might be observed at energies to be reached by the
next generation of colliders \cite{Frampton}. In the \tto, leptons have
the following representation under the \suu~ gauge group

\begin{equation}
\begin{array}{cc}
\left(\begin{array}{c} 
e\\
\nu_e\\
e^c
\end{array}\right),
\left(\begin{array}{c} 
\mu\\
\nu_{\mu}\\
\mu^c
\end{array}\right),
\left(\begin{array}{c} 
\tau\\
\nu_{\tau}\\
\tau^c
\end{array}\right)  
\end{array}:~(1,3^*,0). 
\end{equation}

\noindent In the quark sector, the first two quark families are
represented alike

\begin{equation} \begin{array}{c}

\begin{array}{cc}
\left(\begin{array}{c} 
u^{\alpha}\\
d^{\alpha}\\
D^{\alpha}
\end{array}\right),
\left(\begin{array}{c} 
c^{\alpha}\\
s^{\alpha}\\
S^{\alpha}
\end{array}\right)
\end{array}
:~(3,3,-1/3),\\ \\
\begin{array}{cc}
\begin{array}{c}
u_{\alpha}^c,\\
d_{\alpha}^c,\\
D_{\alpha}^c,
\end{array}
\begin{array}{c}
c_{\alpha}^c:~(3,1,-2/3),\\
s_{\alpha}^c:~(3,1,+1/3),\\
S_{\alpha}^c:~(3,1,+4/3),
\end{array}
\end{array}

\end{array}
\end{equation}

\noindent whereas the third quark family, which is treated differently, is
represented by

\begin{equation}
\begin{array}{c}

\left(\begin{array}{c} 
b^{\alpha}\\
t^{\alpha}\\
T^{\alpha}
\end{array}\right)

:~(3,3^*,2/3),\\ \\

\begin{array}{c}
b_{\alpha}^c:~(3,1,+1/3),\\
t_{\alpha}^c:~(3,1,-2/3),\\
T_{\alpha}^c:~(3,1,-5/3), 
\end{array}

\end{array}
\end{equation}

\noindent with $\alpha=1,2,3$ the quark color number. The charge operator
is defined by $Q/e=\lambda^3/2+\sqrt{3} \lambda^8/2+X$, where
$\lambda^{3,8}$ are the usual Gell-Mann matrices. In this way, the new
quarks $D$, $S$, and $T$ have charge $-4/3~e$, $-4/3~e$, and $5/3~e$,
respectively. In the gauge sector there are five extra vector bosons
besides those of the SM: a pair of singly charged gauge bosons $U^{\pm}$,
a pair of doubly charged ones $U^{\pm \pm}$, and a neutral one
$Z^{\prime}$. In the minimal version of the \tto, three Higgs triplets
$\eta$, $\rho$, and $\chi$ as well as one Higgs sextet $\xi$ are required
in order to accomplish symmetry breaking. The exotic quarks and the five
new gauge bosons acquire mass when the triplet $\eta$ breaks \suu ~ down
to the electroweak gauge group. The last stage in symmetry breaking occurs
when the extra Higgs multiplets $\rho$, $\chi$ and $\xi$ break the
electroweak gauge group down to $\mathrm{U_{e}}$.

\section{Numerical results and discussion} 

It is evident that the most distinctive signal of the \tto~ might be given
by the doubly charged bilepton and the exotic quarks. Studies derived from
low energy data show that the mass of the doubly charged bosons and that
of the exotic quarks might lie below the TeV scale. In particular, from
constraints on the electroweak parameters it was found that 230 GeV $\leq
M_{U^{\pm \pm}} \leq$ 800 GeV \cite{Frampton}, whereas LEP searches for
SUSY particles give the bound $M_{D,S,T} \geq$ 250 GeV \cite{Das}.
Recently, the more stringent bound $M_{U^{\pm \pm}} \geq$ 850 GeV was
derived from muonium-antimuonuim conversion, which would rule out the
minimal \tto ~ but not those versions with an extended Higgs sector
\cite{Willman}. In fact, it has been argued that less stringent bounds are
not excluded since in obtaining such constraint it was assumed that the
couplings of bileptons to leptons are given by the identity matrix, which
is a very restrictive condition indeed\cite{Pleitez}. In this letter we
will consider the range 0.3--1 TeV for the mass of both the doubly charged
bileptons and the exotic quarks, because it is this range which would be
of interest at a future LC.

\begin{figure}[ht]
\centerline{\epsfig{file=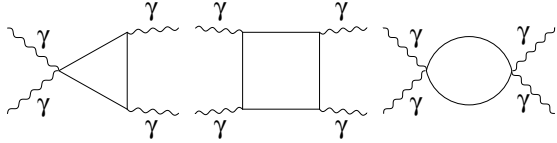,width=7.5cm }}
\caption{Generic Feynman diagrams for $\gamma \gamma \to \gamma \gamma$
scattering in the nonlinear $R_{\xi}$ gauge. Charged particles of spin 0 and 1
contribute through all these diagrams, whereas fermions
participate just via the box diagram.} 
\label{fig:0}
\end{figure}

The amplitude of light by light scattering receives contributions from the
Feynman diagrams shown in Fig. \ref{fig:0}. In addition to the SM
contribution, any extended model produce virtual effects on this process
through loops carrying charged particles. \footnote{We are not considering
extra dimension theories, where light by light scattering proceeds through
the exchange of spin-2 gravitons coupling to photons at tree level.} The
new contributions depend exclusively on the mass and electric charge of
the non standard particles, a feature which allows us to examine in a
model independent way the virtual effects arising from a certain class of
charged particles, namely scalars, vectors or fermions. Those new physics
effects would manifest as deviations from the SM cross section, for this
reason light by light scattering turns out to be useful: the invariant
amplitude contributed by the new particle is proportional to the electric
charge factor $Q^4$, which becomes $Q^8$ when the amplitude is squared to
calculate the cross section. As a result, the total cross section is
significantly enhanced in models including particles with a charge whose
absolute value is greater than unity, in terms of the positron charge.
Moreover, the enhancement factor $Q^8$ is very useful to distinguish
clearly between the contributions coming from particles with the same spin
and mass but a different electric charge. For instance, the contribution
from a doubly charged particle is a factor of $2^8=256$ larger than that
from a singly charged particle of the same class.

\begin{figure}[ht]
\centerline{\epsfig{file=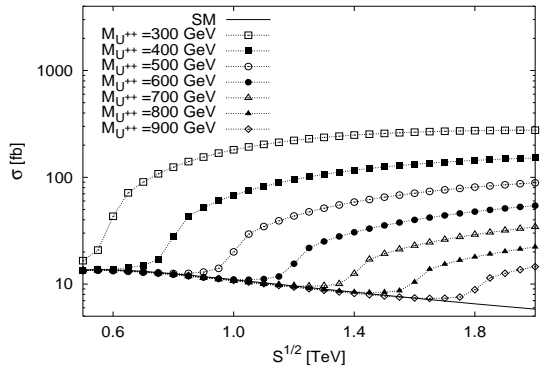,width=7.5cm,height=5cm }}
\caption{Unpolarized cross section for $\gamma \gamma \to
\gamma \gamma$ scattering when it is added to the SM a
doubly charged vector bilepton $U^{++}$.} 
\label{fig:1}
\end{figure}

\begin{figure}[ht]
\centerline{\epsfig{file=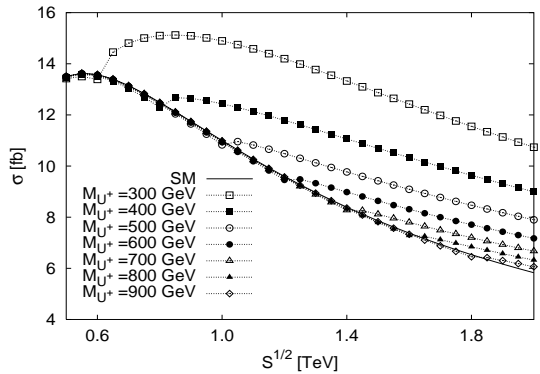,width=7.5cm,height=5cm }}
\caption{The same as in Fig. \ref{fig:1} for the case of 
a singly charged vector bilepton $U^{+}$.} 
\label{fig:2}
\end{figure}

\begin{figure}[ht]
\centerline{\epsfig{file=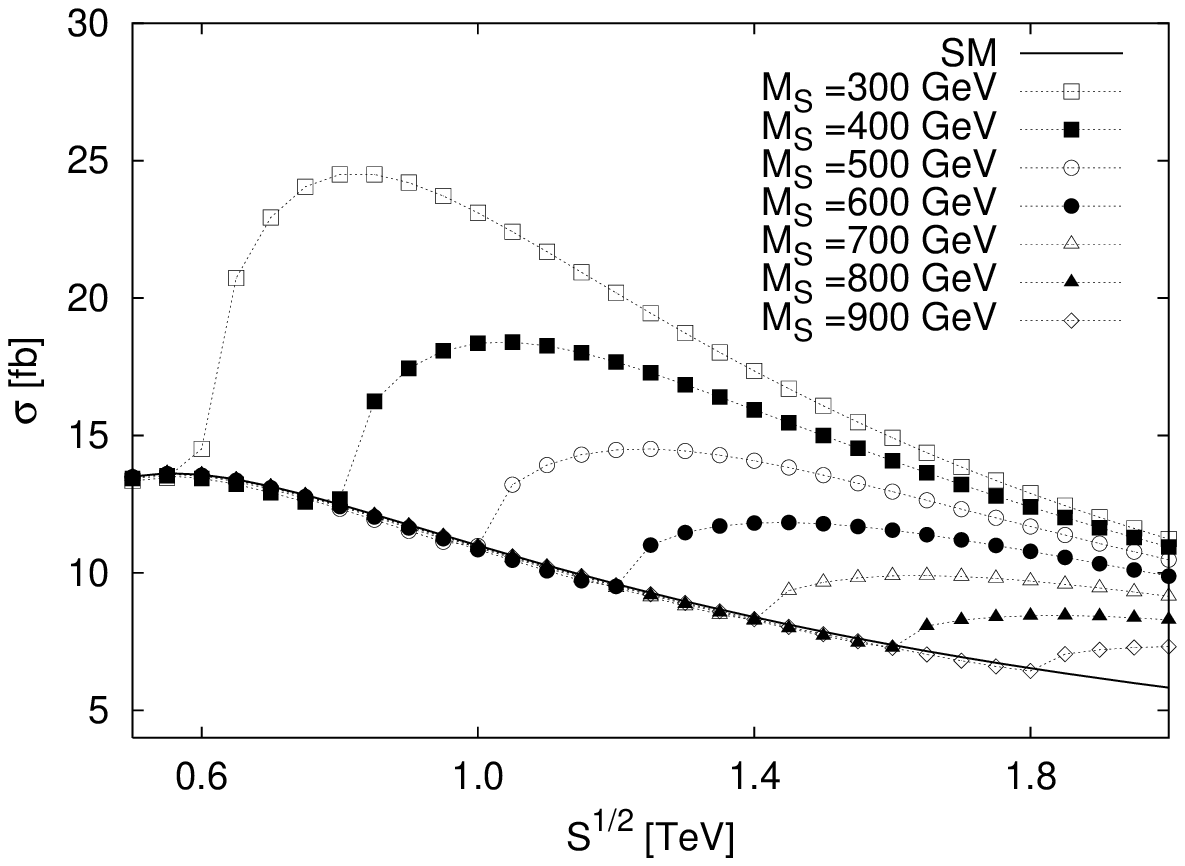,width=7.5cm,height=5cm }}
\caption{The same as in Fig. \ref{fig:1} for the case 
of an exotic quark $S$ with charge $5/3~e$.} 
\label{fig:3}
\end{figure}

\begin{figure}[ht]
\centerline{\epsfig{file=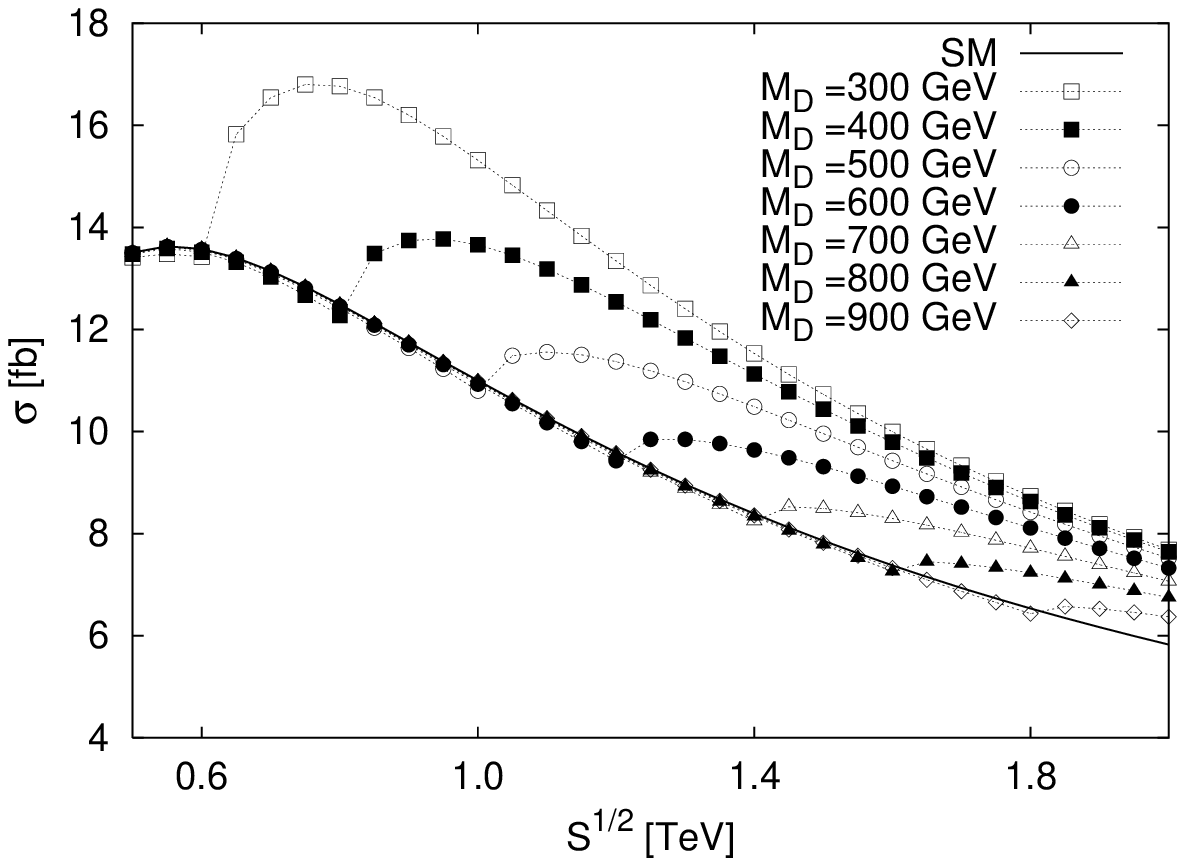,width=7.5cm,height=5cm }}
\caption{The same as in Fig. \ref{fig:1} for the case of 
an exotic quark $D$ with charge $-4/3~e$.} 
\label{fig:4}
\end{figure}

The cross section for $\gamma \gamma \to \gamma \gamma$ scattering can be
obtained by means of the helicity amplitudes for loops with scalar bosons,
fermions or gauge bosons, whose expressions in terms of scalar functions
have been obtained previously \cite{Jikia,Renard}. We have used the
{\small{FF}} numerical routines to evaluate the required scalar functions
\cite{Oldenborgh}. In calculating the cross section, the integration over
the scattering angle $\theta$ has been constrained to lie in the range
$30^{\mathrm{o}} \leq \theta \leq 150^{\mathrm{o}}$, in order to avoid
collinear (soft) photons escaping from the detector. In Figs.
\ref{fig:1}--\ref{fig:4} we have plotted the unpolarized cross sections
which are obtained when each new particle predicted by the \tto~ is
included together with the SM contribution. In all these graphs the
resulting cross section deviates from the SM one at the threshold
$\sqrt{s}\geq 2 M_{\mathrm{New}}$, with $M_{\mathrm{New}}$ the mass of the
new particle. This fact can be understood from the fact that the
additional contribution arises mainly from the interference between the SM
amplitude $A_{\mathrm{SM}}$ and that of the new particle
$A_{\mathrm{New}}$ \cite{Renard}. Therefore, the new cross section will
deviate from the SM one by the term $2 Re(A_{\mathrm{SM}} A_{\mathrm
{New}})$, which is approximately the same as $2 Im(A_{\mathrm{SM}})
Im(A_{\mathrm{New}})$ since at high energies the SM amplitude is almost
purely imaginary. Another interesting fact which is evident in the plots
is that the most distinctive effects come from the doubly charged vector
bilepton, which dominates even to the exotic quarks contribution, although
in the latter case the amplitude gets also enhanced by an additional
factor corresponding to the quark color number. On the other hand, the
less spectacular effects arise from a singly charged bilepton gauge boson,
which together with a doubly charged scalar bilepton have been previously
studied in the context of LRM and SUSYLR \cite{Toscano1}.

The viability of using light by light scattering to search for SUSY
particles has been examined with detail in \cite{Renard}. It was shown
that under the conditions expected at the planned LC, the unpolarized
cross section would be sensitive enough to allow the detection of
contributions coming from charginos and sleptons. It is clear that, due to
the enhancement produced by their charge, the detection of the exotic
particles predicted by the \tto ~ is more promising than that of SUSY
particles. In this respect, in searching for exotic particles at the
present and future colliders, direct production represents also a viable
mode. However, any theoretical analysis done in this direction has the
disadvantage of requiring many assumptions about unknown model-dependent
parameters. For instance, if direct production of bileptons is analyzed,
one must consider parameters such as their couplings to leptons, the mass
of the extra neutral boson $M_{Z^{\prime}}$, and the $Z-Z^{\prime}$ mixing
angle \cite{Cuypers}. This unwanted situation also arises in considering
indirect search modes such as M{\o}ller and Bhabha scattering, which at
tree level receive the contribution of a doubly charged bilepton through
the $u$ channel. Furthermore, an experimental study of direct production
must consider also the respective backgrounds. In contrast, the cross
section of light by light scattering involves just one free parameter,
which is the mass of the exotic particle. Although the cross sections are
smaller than the ones predicted in another reactions, the facilities which
would be offered at a LC open up the possibility of using light by light
scattering as an alternative method to search for non standard particles
such as bileptons and exotic quarks.

\begin{figure}[ht]
\centerline{\epsfig{file=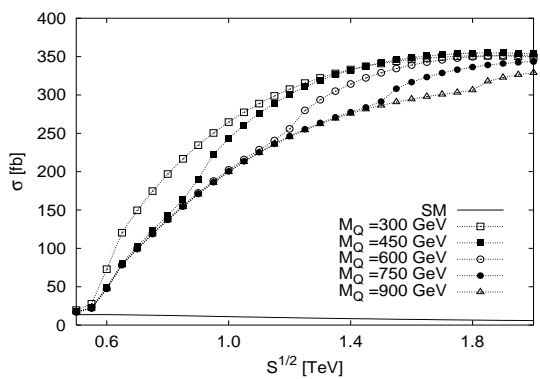,width=7.5cm,height=5cm }}
\caption{Unpolarized cross section for $\gamma \gamma \to \gamma \gamma$
scattering in the \tto. We have assumed a simplistic scenario where the
mass of the three quarks is $M_Q$. Both vector bileptons are given a mass
of $M_{U}=300$ GeV.}
\label{fig:5}
\end{figure}

\begin{figure}[ht]
\centerline{\epsfig{file=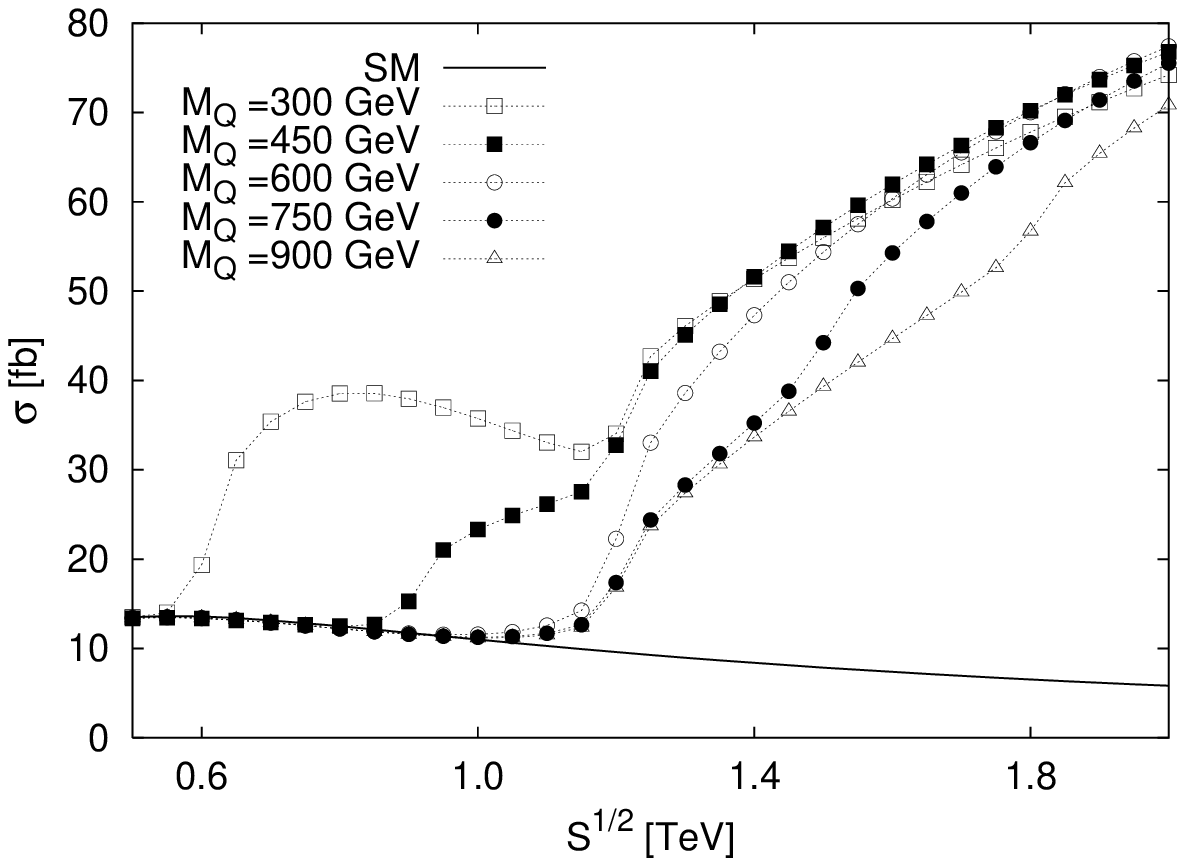,width=7.5cm,height=5cm }}
\caption{The same as in Fig. \ref{fig:5} when the vector bilepton mass is
$M_{U}=600$ GeV.}
\label{fig:6}
\end{figure}

\begin{figure}[ht]
\centerline{\epsfig{file=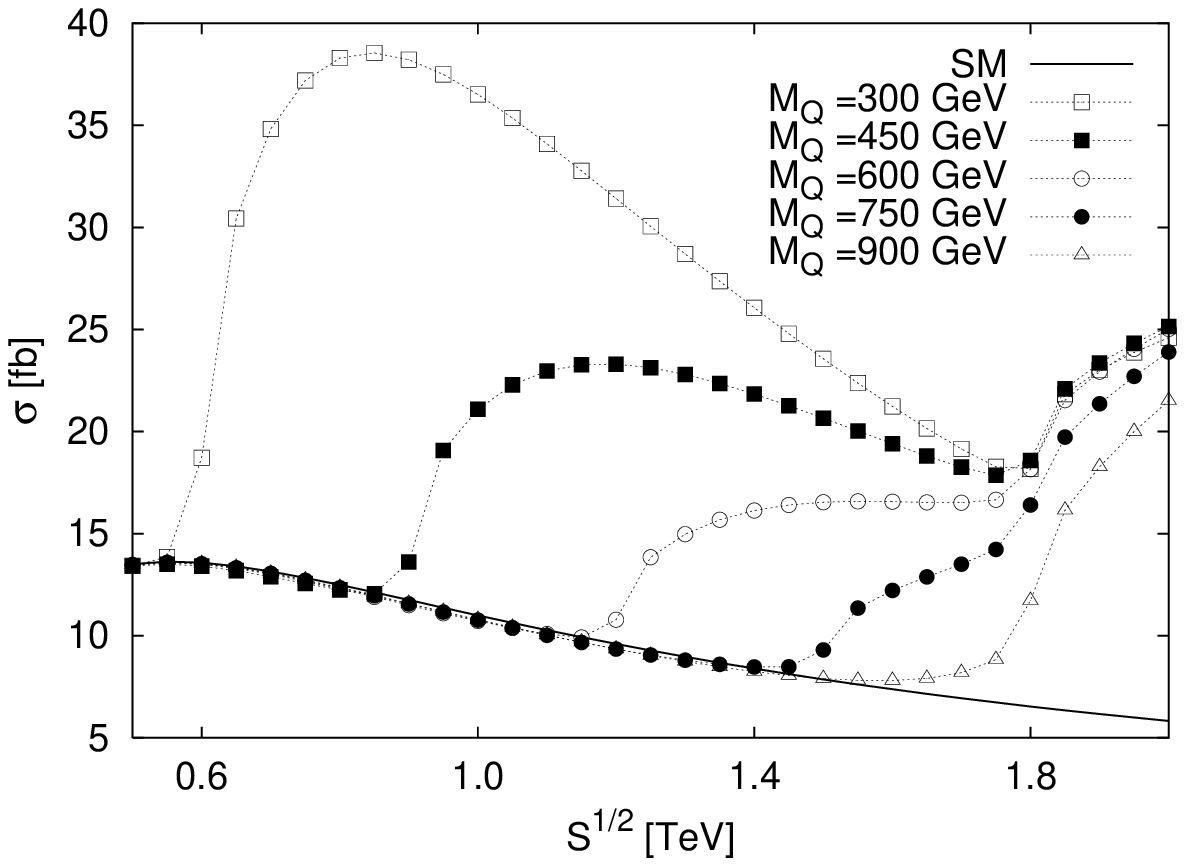,width=7.5cm,height=5cm }}
\caption{The same as in Fig. \ref{fig:5} when the vector bilepton mass is
$M_{U}=900$ GeV.}
\label{fig:7}
\end{figure}

Finally, by way of illustration, we show through Figs
\ref{fig:5}--\ref{fig:7} the total $\gamma \gamma \to \gamma \gamma$
unpolarized cross section for some scenarios which may arise in the \tto.
We have assumed the simple case in which the three exotic quarks have a
mass $M_Q$ and both vector bileptons have a mass $M_U$. Such scenarios are
very simplistic, but we only want to show how spectacular is the
enhancement produced by the full particle content of the \tto. As shown in
Fig. \ref{fig:5}, the most notable scenario is the one with a relatively
light doubly charged bilepton with a mass of a few hundreds of GeVs. In
such a case the cross section may be one order of magnitude larger than
the SM one. Although the doubly charged bilepton might produce spectacular
effects by itself in this process and in any other ones, its detection
would not be a sufficient evidence to support the \tto, and further probes
would be needed. In this respect, a reaction as light by light scattering
might be an useful mode to test indirectly some details of the model. As
the cross section is enhanced considerably by both bileptons and exotic
quarks, the experimental study of light by light scattering would produce
indirect evidence of the existence of these particles. In addition, since
all the new contributions add up coherently, this mode would be suitable
for counting the number of new particles, what would be helpful to
identify a particular model. However, it is important to note that the
scope of using light by light scattering to search for new particles is
limited by the fact that any virtual effect appear at the threshold which
corresponds to twice the mass value of the new particle.

\section{Summary}

In closing, we stress that light by light scattering offers an interesting
mode to search indirectly for exotic particles, especially the ones
predicted by the \tto, at a future LC. In particular, doubly charged
vector bileptons give rise to spectacular new physics effects. If LC
became a reality, light by light scattering would be useful to elucidate
what is the adequate gauge group to extend the SM.

\acknowledgments

We acknowledge support from CONACYT and SNI (M\'exico).

\end{document}